\documentclass[aps, prl, twocolumn, amsfonts, amsmath]{revtex4-1}
\usepackage{graphicx}
\usepackage{amssymb,amsmath}
\usepackage{soul}
\usepackage{natbib}

\renewcommand{\vec}[1]{{\mathbf #1}}

\usepackage[usenames,dvipsnames]{color}

\begin{document}
\title{Correlations between reflected and transmitted intensity patterns emerging from opaque disordered media}

\author{I. Starshynov,$^{1, \ast}$ A. M. Paniagua-Diaz,$^{1, \ast}$ N. Fayard, $^{2, \ast}$ A. Goetschy,$^{2}$ R. Pierrat,$^{2}$ R. Carminati,$^{2}$ J. Bertolotti$^{1}$}
\affiliation{$^{1}$University of Exeter, Stocker Road, Exeter EX4 4QL, United Kingdom\\
$^{2}$ESPCI Paris, PSL Research University, CNRS, Institut Langevin, 1 rue Jussieu, F-75005, Paris, France\\
$^\ast$These authors contributed equally to this work.
}
\maketitle

{\bfseries 
The propagation of monochromatic light through a scattering medium produces speckle patterns in reflection and transmission, and the apparent randomness of these patterns prevents direct imaging through thick turbid media. Yet, since elastic multiple scattering is fundamentally a linear and deterministic process, information is not lost but distributed among many degrees of freedom that can be resolved and manipulated. Here we demonstrate experimentally that the reflected and transmitted speckle patterns are correlated, even for opaque media with thickness much larger than the transport mean free path, proving that information survives the multiple scattering process and can be recovered. The existence of mutual information between the two sides of a scattering medium opens up new possibilities for the control of transmitted light without any feedback from the target side, but using only information gathered from the reflected speckle.
}

In multiply scattering materials, the random inhomogeneities in the refractive index scramble the incident wavefront, mixing colors and spatial degrees of freedom, resulting in a white and opaque appearance~\cite{Sebbah1999aa}. Under illumination with coherent light and for elastic scattering, interference produces large intensity fluctuations that is not averaged out by a single realization of the disorder, resulting in a seemingly random speckle pattern~\cite{shapiro86}. In principle the speckle pattern encodes all the information on the sample and the incident light~\cite{Freund1990}. A complete knowledge of the scattering matrix allows one to reverse the multiple scattering process and to recover the initial wavefront, thus permitting imaging through turbid materials~\cite{Popoff2010aa,mosk12}. Conversely, if the scattering matrix is not known, a multiply scattering material effectively behaves as an opaque screen.

Speckle patterns are not as random as they appear at first sight. Interference between the possible scattering paths in the medium are known to produce spatial correlations between the intensity measured at different positions~\cite{berkovits94, vanrossum99,akkermans07}, and correlations of different ranges have been identified~\cite{Feng1988aa}. Short-range correlations determine the size of a speckle spot. Long-range correlations emerge as a consequence of constraints such as energy conservation or reciprocity~\cite{genack90, deboer92, scheffold98}. Spatial correlations have not been used for imaging, a notable exception being the optical memory effect~\cite{freund88}, a correlation of purely geometrical origin that has been exploited for non-invasive imaging through an opaque scattering layer~\cite{bertolotti12, katz14}.

At first glance, as transmitted and reflected waves are expected to undergo very different multiple scattering sequences, correlations between transmitted and reflected wavefronts are expected to quickly average to zero. Very little attention has been given to cross-correlations between transmitted and reflected speckles, their existence being only mentionned in passing~\cite{rogozkin95, froufe07}.  However, a recent theoretical study suggested that a long-range correlation should survive even for thick (opaque) scattering media~\cite{fayard15}. The existence of this reflection-transmission correlation suggests that one could non-invasively extract information on the transmitted speckle from a measurement restricted to the reflection half-space. 

Here we report the first measurement of the intensity correlation between transmitted and reflected speckle patterns, for scattering materials with thickness $L$ and scattering mean free path $\ell$ covering all the range from single scattering ($L\lesssim\ell$) to diffusive transport ($L \gg \ell$). The data are supported by 3D numerical simulations, and by a theoretical analysis of the lineshape of the correlation function, and its dependence on the experimental parameters. The experiments and the theory embrace the complexity and the richness of the phenomenon, thus opening the way to its use as a basic ingredient in the design of new approaches for sensing, imaging or communicating through opaque scattering media.

\begin{figure}[bt]
\includegraphics[width=0.95\linewidth]{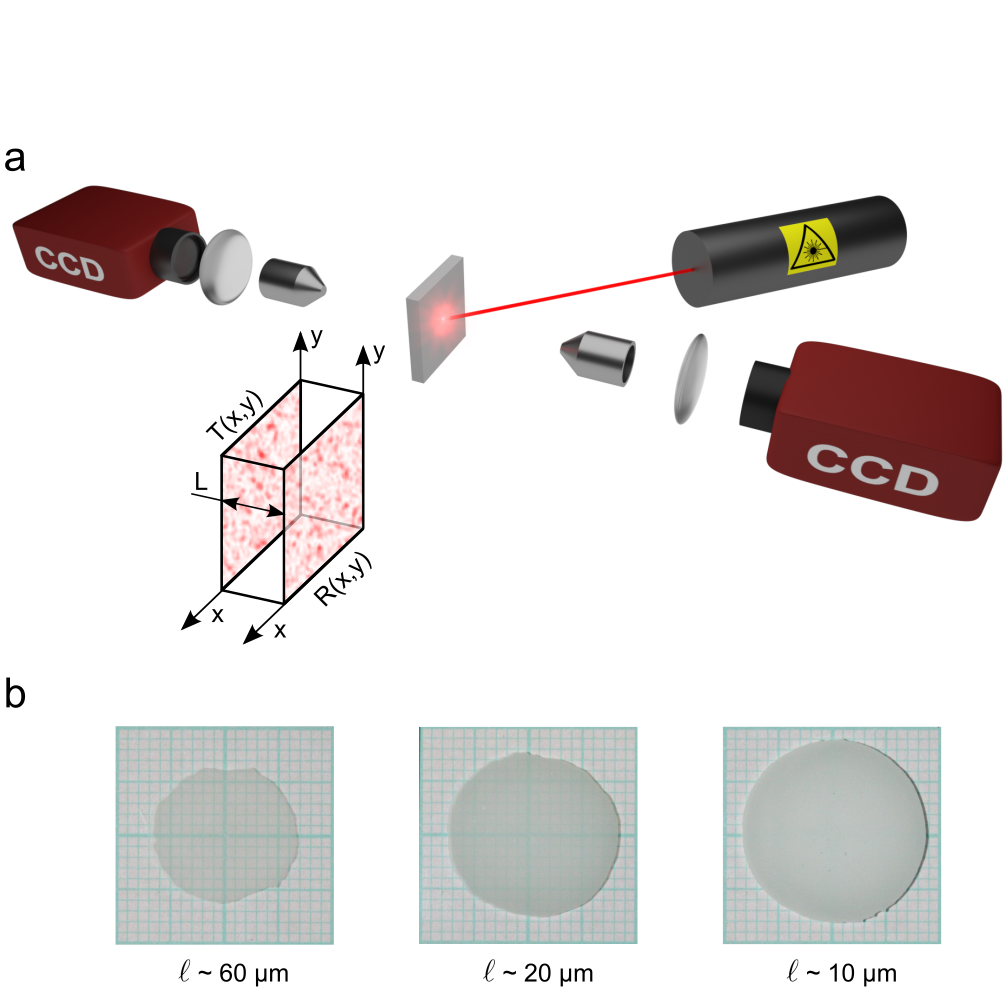}
\caption{(a) Experimental setup. A scattering slab, formed by a suspension of TiO$_2$ particles in glycerol, is illuminated by a laser beam incident at an angle $\sim45^{\circ}$.  The speckle patterns on the two surfaces, $T(x,y)$  and $R(x,y)$ respectively, are recorded with two identical imaging systems. (b) Examples of samples with thickness $L=20\mu$m but different TiO$_2$ concentrations: from left to right 5 g/l, 10 
g/l and 40 g/l, which correspond to a mean free path of (60, 20.4 and 9.8) $\pm$ 2.5 $\mu$m, respectively. }
\label{fig:schematic}
\end{figure}

\begin{figure}[tb]
\includegraphics[width=0.95\linewidth]{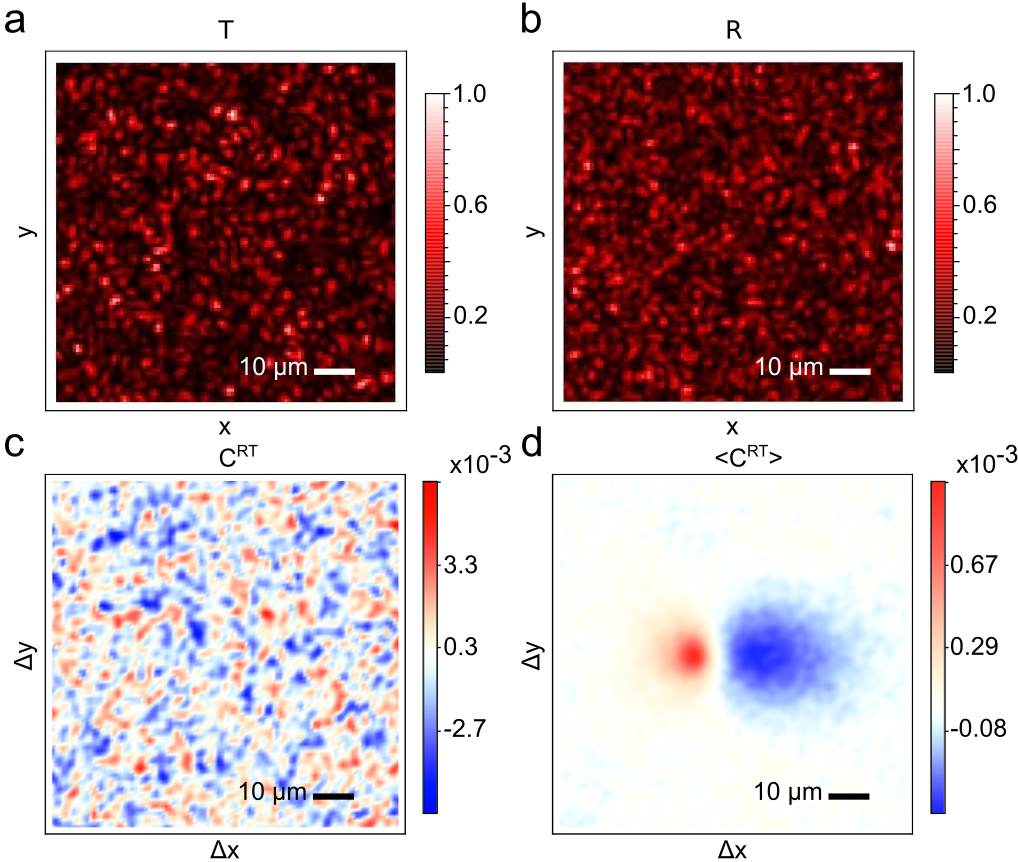}
\caption{Typical measured speckle patterns in transmission (a) and reflection (b), for a sample with $L= 20\, \mu$m and $\ell \simeq 60\, \mu$m. (c) Correlation function $C^{RT}(\Delta x,\Delta y)$ computed from the speckle patterns in (a) and (b). (d) Ensemble averaged correlation function $\langle C^{RT} (\Delta x, \Delta y)\rangle$  obtained from 10$^4$ realizations of the disorder. The long-range character of the correlation function, that extends far beyond the size of a speckle spot, is clearly visible.} 
\label{fig:speckles}
\end{figure}

The experimental apparatus is shown in Fig.~\ref{fig:schematic}a. A monochromatic wave is incident at an angle $\sim45^{\circ}$ on a suspension of TiO$_2$ particles in glycerol, held between two microscope slides to form a scattering slab. The slab thickness $L$ is controlled using calibrated spacers, and the mean free path $\ell$ is controlled by varying the TiO$_2$ concentration. Typical samples with different optical thickness $b=L/\ell$, from semitransparent to fully opaque are shown in Fig.~\ref{fig:schematic}b. For a set of given $L$ and $\ell$ we record the intensity patterns $R(\vec{r})$ and $T(\vec{r})$ on the surface of the sample in reflection and transmission, respectively, with two identical imaging systems (see Methods). As the samples are liquid the resulting speckle patterns change in time due to Brownian motion of the scatterers, with a correlation time $\tau$ that depends strongly on the sample thickness.  Choosing an integration time $<\tau$, and a time interval between successive measurements $>\tau$, allows us to measure speckle images $R(\vec{r})$ and $T(\vec{r})$ for a large ensemble of configurations of the disordered medium. An example pair of images measured for a given realization of disorder is shown in Fig.~\ref{fig:speckles}a,b. For each pair of $R$ and $T$ we calculate the correlation function $C^{RT}$ defined as
\begin{equation}
\label{eq:Corr}
C^{RT}(\Delta \vec{r}) =  \frac{\overline{\delta R(\vec{r}) \delta T(\vec{r}+ \Delta \vec{r})}}{\left(\overline{ [\delta R(\vec{r}) - \overline{\delta R(\vec{r})}]^2} \cdot \overline{ [\delta T(\vec{r}) - \overline{\delta T(\vec{r})}]^2}\right)^{1/2} }
\end{equation}
where $\Delta \vec{r}=(\Delta x,\Delta y)$ is a transverse shift between the images, and the overline denotes the spatial average over the coordinates $\vec{r}$. We have introduced the notation $\delta f = f - \langle f \rangle$ for the statistical fluctuations of a random variable $f$, with $\langle \cdot \rangle$ representing the ensemble average over disorder. Plotted as a 2D map the correlation function $C^{RT}(\Delta x, \Delta y)$ appears random, with a granularity similar to that of a speckle image (Fig.~\ref{fig:speckles}c). After ensemble averaging over the realizations of the disorder a clear pattern emerges in
 $\langle C^{RT} (\Delta x, \Delta y) \rangle$ (Fig.~\ref{fig:speckles}d), showing that the transmitted and reflected speckle patterns are indeed correlated.

Speckle correlations are commonly divided into three categories: Short-range correlations ($C_1$) that decay with the separation between the observation points on the scale of the wavelength, long-range correlations ($C_2$) that have a polynomial decay, and infinite-range correlations ($C_3$)~\cite{Feng1988aa,akkermans07}. The short-range correlation $C_1$ corresponds to the approximation of a field obeying Gaussian statistics~\cite{shapiro86}, while $C_2$ and $C_3$ are non-Gaussian corrections. An additionnal infinite-range correlation ($C_0$) appears under illumination by a point source located inside the medium~\cite{shapiro99}. One can see in Fig.~\ref{fig:speckles}d that the lineshape of $\langle C^{RT} (\Delta \vec{r})\rangle$ is much wider than a speckle spot, indicating that the dominant contribution to this correlation is long-range in nature.

\begin{figure*}[t!h]
{\includegraphics[width=0.95\linewidth]{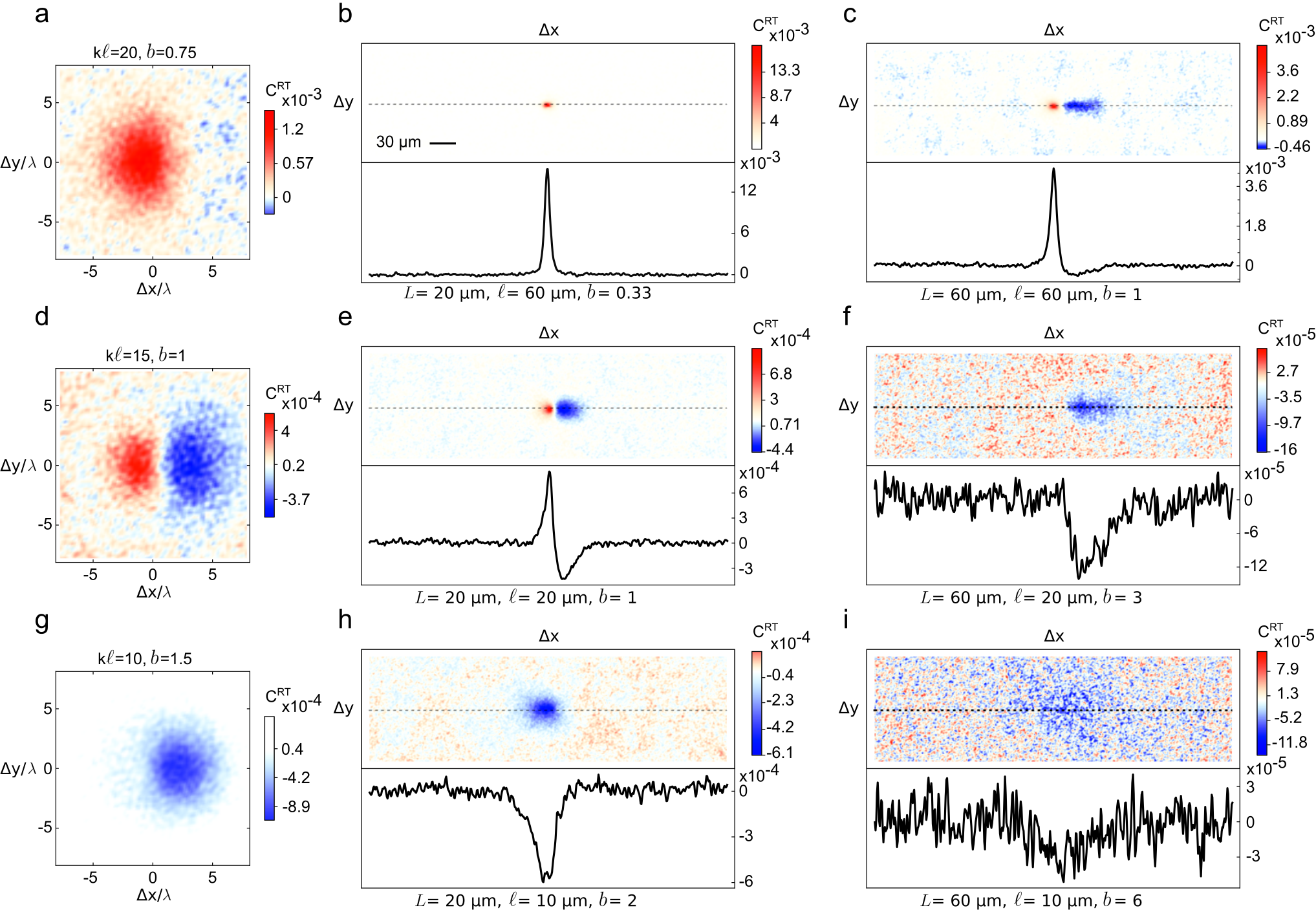}}
\caption{Average reflection-transmission correlation function $\langle C^{RT} \rangle$ for different values of $L$ and $\ell$ and the optical thickness $b=L/\ell$. Left column: 3D numerical simulations of 2D maps of $\langle C^{RT} (\Delta x, \Delta y)\rangle$. Center and right colums: Experimental results. For clarity, both 2D maps of $\langle C^{RT} (\Delta x, \Delta y)\rangle$ and cross-sections along the line $\Delta y =0$ (indicated as a dotted line in the 2D maps) are displayed. Two regimes are identified. For moderate optical thickness ($b\lesssim 1$), the correlation function is dominated by a narrow peak with a negative side lobe. For large optical thicknesses, ($b > 1$), the correlation function is dominated by a wide negative dip.}
	\label{fig:result}
\end{figure*}

In order to characterize the lineshape of the correlation function, and to probe its dependence on the sample parameters, we measured $\langle C^{RT} (\Delta\vec{r})\rangle$ for different values of $\ell$ and $L$, covering  the full range from the single scattering ($L\lesssim \ell$) to the diffusive ($L \gg \ell$) regime. The results are summarized in Fig.~\ref{fig:result} (center and right columns), where both 2D maps $\langle C^{RT} (\Delta x, \Delta y)\rangle$ and cross-sections along the line $\Delta y =0$ (indicated as a dotted line in the 2D maps) are displayed. It is interesting to note that both the shape and the sign of the reflection/transmission correlation substantially depend on $L$ and  $\ell$. In the single scattering regime (optical thickness $b\lesssim 1$),  $\langle C^{RT} \rangle$ is dominated by a narrow peak (still much larger than a single speckle spot) with a negative side lobe. In the multiple scattering regime ($b \gg 1$), $\langle C^{RT} \rangle$ is dominated by a wide negative dip.

The short-range contributions to $C^{RT}$ ($C_1$) decay on the scale of the wavelength~\cite{shapiro86}, and are thus negligible in all measurements since in the reflection-transmission geometry the observation points are separated by a distance $\sqrt{L^2+|\Delta \vec{r}|^2}$ which is much larger than $\lambda$ for even the thinniest sample (see SI section 2). Hence, $\langle C^{RT} \rangle$ is necessarily a long-range correlation of the $C_2$ type. It is interesting to note that the reflection/transmission geometry naturally favors the observation of non-Gaussian long-range correaltions, without requiring any post-processing to remove the $C_1$ contribution that dominates in the pure transmission geometry~\cite{genack90, deboer92, sebbah00, strudley13}.
Another feature of our experiment is the illumination/detection geometry that excludes any contribution from specularly reflected and transmitted averaged fields (see Methods). Indeed, in the geometry in Fig.~\ref{fig:schematic}a, the averaged field is not collected by the detectors, and we directly correlate $T(\vec{r})=\vert \delta E_T(\vec{r}) \vert^2$ and $R(\vec{r})=\vert \delta E_R(\vec{r}) \vert^2$, thus avoiding spurious interference terms in the correlation function for the single scattering regime (see SI section 3). 

To support the experimental data, we have performed full numerical simulations of wave propagation in three-dimensional disordered media. In the simulations, the samples consist of slabs of dipole scatterers with random positions. The scalar wave equation is solved numerically using the coupled-dipole method (see Methods). From the computation of the complex amplitude of the scattered field in reflection and transmission, we deduce $\langle C^{RT} \rangle$ performing the ensemble average over a large set of samples. The results of the simulations are displayed in Figs.~\ref{fig:result} (left column), and are in very good agreement with the experimental data. The general shape of the correlation in the regimes $b<1$, $b \simeq 1$ and $b >1$ is well reproduced in the simulations.

In order to refine the analysis, and to get more physical insight, we have also used a formal pertubation theory to calculate the shape of the correlation function $\langle C^{RT} \rangle$. In this formal approach, the correlation function $\langle C^{RT}(\Delta \vec{r}) \rangle =\langle \delta R(\vec{r}) \delta T(\vec{r}+ \Delta \vec{r}) \rangle/\langle R(\vec{r}) \rangle \langle T(\vec{r}) \rangle$ is directly computed from a statistical ensemble averaging, without going through the intermediate spatial average in Eq.~(\ref{eq:Corr}) used for the experimental data. Both averaging processes coincide provided that $\ell \gg \lambda$, a condition that is always satisfied in our experiments (see Methods and SI section 1). Formal pertubation theory uses $1/(k \ell)$ as a small parameter, with $k=2\pi/\lambda$, and relies on a diagrammatic formalism that allows one to derive explicit expressions of intensity correlation functions~~\cite{berkovits94, vanrossum99,akkermans07}. In the reflection/transmission geometry, care must be taken to properly account for leading contributions~\cite{rogozkin95, rogozkin96}. 

\begin{figure}[th]
	{\includegraphics[width=1\linewidth]{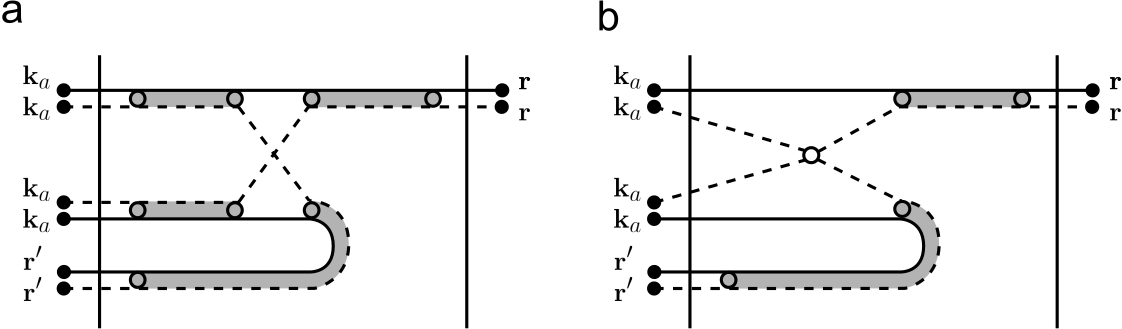}}
	\caption{Diagrams contributing to the $\langle C^{RT} (\Delta\vec{r})\rangle$ correlation. An intensity correlation depends on 2 intensities (4 fields) that propagate through the sample, therefore involving 4 inputs and 4 outputs. Shaded tubes represent diffusive paths and open circles stand for scatterers; single solid lines stand for averaged fields and single dashed lines for their complex conjugates. The diagram in panel (a) is representative of the class of $C_2$ diagrams describing the negative contribution of the correlation function at large optical thickness ($L \gg \ell$). Panel (b) represents the class of $C_0$-type diagrams that contribute to the positive peak dominant in the regime $\ell \sim L \gg \lambda$.}
	\label{fig:diagrams}
\end{figure}

Let us first discuss the regime of large optical thickness $L\gg \ell$ corresponding to Fig.~\ref{fig:result}f-i. Strikingly, we observe in this regime that $\langle C^{RT} (\Delta\vec{r})\rangle$ is negative, in agreement with the prediction in Ref.~\cite{fayard15}. This means that for every bright spot in reflection (transmission) the corresponding area in  transmission (reflection) is more likely to be darker, and {\it vice versa}. This feature can be inferred from flux conservation arguments. Indeed, defining $T\varpropto \int T (\vec{r}) \textrm{d}\vec{r}$ and $R\varpropto \int R (\vec{r}) \textrm{d}\vec{r}$, energy conservation imposes $T+R=1$ for a non-absorbing medium, from which we can show that $\int \langle C^{RT} (\Delta\vec{r})\rangle \textrm{d} \Delta\vec{r} \varpropto \langle \delta T \delta R \rangle=- \langle \delta T^2\rangle <0$ (see SI section 4). Note that the existence of negative long-range $C_2$ correlations has been previously pointed out in Refs.~\cite{saenz03, froufe07}. Refining the analysis performed in Ref.~\cite{fayard15} (see SI section 5), we have studied theoretically the regime $L\gtrsim \ell \gg \lambda $, and found that both the amplitude and the width of the correlation function depend on $L$ and  $\ell$, as in the experimental data in Figs.~\ref{fig:result}f-i. For $L \gg \ell$, the dominant diagrams belong to the class represented in Fig.~\ref{fig:diagrams}a, that are typical of long-range $C_2$ correlation functions. They predict a correlation function that is isotropic, independent of the angle of incidence, and scales as $\langle C^{RT} (\Delta\vec{r})\rangle = C_2^{RT} (\Delta\vec{r}) = -f(|\Delta\vec{r}|)/(kL)^2 $, where $f$ is a dimensionless function that decays on a range $|\Delta\vec{r}|\simeq L$~\cite{fayard15}. This long-range character of the correlation function originates from the crossing of two diffusive paths that probe a transverse distance $L$, as represented in Fig.~\ref{fig:diagrams}a. Moreover, the correlation function in this regime is independent of the disorder strength $k\ell$, which makes it strikingly different  from that observed in a pure transmission geometry, for which $C_2^{TT} \sim 1/[(k\ell)(kL)] \varpropto 1/g$, where $g$ is the dimensionless conductance of the sample~\cite{feng88}.  Another important difference between $C_2^{RT}$ and $ C_2^{TT} $ is the evolution of their information content with respect to the detection scheme. 
Although $C_2^{TT}$ contains the same information whether it is measured on the sample surface or in the far field, this is not the case for $C_2^{RT}$. 
Indeed, in the far field, we have $\langle C^{RT}(\vec{k}_b, \vec{k}_{b'})\rangle\sim  \int \langle C^{RT} (\Delta\vec{r})\rangle \textrm{d} \Delta\vec{r} = \text{const.}$ for any pair of observation directions $\vec{k}_b, \vec{k}_{b'}$, as the information content is spread uniformly over all degrees of freedom. Therefore we will focus our discussion on the correlations measured on the sample's surface.

In the regime of moderate optical thickness $\ell \sim L \gg \lambda$, where single scattering is expected to dominate, an intensity correlation extending far beyond the size of a single speckle spot is still observed (see Fig.~\ref{fig:result}c-e), but with a positive peak appearing in the vicinity of the negative contribution. The apparent relative position and amplitude between the peak and the dip depends on the angle of incidence of the illumination (see SI section 6). Contrary to the negative dip in the correlation function observed at large optical thickness, the lineshape is anisotropic, with negative side lobes (hardly visible in Fig.~\ref{fig:result}c-e but visible in a calculation shown in SI section 6) that are more pronounced along the direction of the incidence plane. Moreover, the amplitudes of both the positive peak and the side lobes substantially depend on the incidence angle. These two features of the correlation function in the regime $\ell \sim L \gg \lambda$ (long-range extent and dependence on the angle of incidence) suggest a qualitative description based on diagrams of the class represented in Fig.~\ref{fig:diagrams}b, that satisfy both properties simultaneously (see SI section 7). Computing these diagrams lead to a contribution to the correlation function scaling as $1/(kL)^4$ for $b\gg 1$. This is consistent with the fact that, according to the measurements and the numerical simulations, this contribution has to be negligible at large optical thickness, where the $C_2^{RT}$ correlation function discussed previously scaling as $1/(kL)^2$ dominates.
The observed anisotropy in the correlation function is also well reproduced, supporting the relevance of the analysis based on the diagrams in Fig.~\ref{fig:diagrams}b. Interestingly, these diagrams are formally similar to those leading to the infinite-range correlations $C_0$ observed when the sample is excited by a point source~\cite{shapiro99, hildebrand14}. In a non-absorbing medium, as a consequence of energy conservation, the $C_0$ contribution is related to the fluctuations of the local density of states at the source position~\cite{skipetrov00, caze10}. In the present context, where a plane wave excitation is used, the $C_0$-type contribution to the reflection-transmission correlation function is long-range and satisfies $\int \langle C_0^{RT} (\Delta\vec{r})\rangle \textrm{d} \Delta\vec{r}=0$ (see SI section 7). This property also leads to the conclusion that the $C_0$-type contribution observed is specific to speckle patterns measured on the surface of the sample, and vanishes in the case of far-field angular measurements.

Note finally that in the quasi-ballistic regime $\ell \gg L \gg \lambda$, which is not the focus of our experiment, we expect the correlation $\langle C^{RT} \rangle$ to contain additional contributions to $C_2$ and $C_0$ (see SI section 7), that still result in an overall positive peak.

In summary, we have demonstrated experimentally the existence of a cross-correlation between the speckle patterns measured in reflection and transmission on the surface of a disordered medium. The correlation persists in the regime of large optical thickness $L \gg \ell$ in which the sample is opaque due to multiple scattering. The measurements are supported by 3D numerical simulations, and have been analysed using a perturbative theory (valid when $\ell \gg \lambda$). We have found that the reflection-transmission correlation has two contributions: a positive peak dominant at moderate optical thicknesses $L \lesssim \ell$, and a negative dip dominant in the diffusive regime $L \gg \ell$. The existence of this transmission-reflection correlation proves that mutual information between the two sides of a strongly scattering medium can in principle be exploited, opening new possibilities for the non-invasive control of the transmitted light (or other kind of coherent waves) from measurements restricted to the reflection half-space. This offers new strategies for the detection of objects hidden behind opaque scattering media, including ghost imaging schemes, and the control of wave propagation by wavefront shaping techniques~\cite{ojambati16, hsu17}.

\section{\label{sec:Methods}Methods}
\textbf{Experimental set-up}. 
The experiments are performed using a 2 mW He-Ne laser (632.8 nm model HNLS008L-EC, Thorlabs) incident on the sample at $45^{\circ}$. The imaging system sketched in Fig.~\ref{fig:schematic}a consists of two identical microscope objectives (10X Olympus Plan Achromat Objective, 0.25 NA) and two plano-convex 150 mm lenses. The intensity speckle patterns on the two surfaces of the sample are imaged with two identical CCD cameras (Allied Vision Manta G-146) (see SI section 9). The integration time of the cameras is set to 1 ms, chosen to be shorter than the decorrelation time of the speckle patterns due to Brownian motion of the scatterers. The samples are a mixture of TiO$_2$ particles and glycerol. We prepared the samples with three different TiO$_2$ concentrations: 50, 150 and 400 mg of TiO$_2$  in 10 ml of glycerol, which resulted in a mean free path $\ell$ of (60, 20.4 and 9.8) $\pm$ 2.5 $\mu$m, respectively (see SI section 8). The thickness of each sample is fixed using two feeler gauges of the required thickness. To insure a proper mixing and absence of clustered grains, the TiO$_2$ powder is mixed with glycerol using a magnetic stirrer for two hours, and after that was sonicated for 30 minutes.

\textbf{Definition of the correlation function}. The definition of the correlation function used to analyze the experimental data is given by Eq.~(\ref{eq:Corr}), while in the theoretical analysis we used
\begin{equation}
 \label{eq:theoryC}
 \langle C^{RT}(\Delta \vec{r}) \rangle =\langle \delta R(\vec{r}) \delta T(\vec{r}+ \Delta \vec{r}) \rangle/\langle R(\vec{r}) \rangle \langle T(\vec{r}) \rangle .
\end{equation}The two definitions appear different, but they are rigorously equivalent under two conditions: ergodicity and Gaussian fields. Ergodicity allows us to transform Eq.~(\ref{eq:Corr}) into  $\langle C^{RT} (\Delta\vec{r})\rangle= \langle \delta R(\vec{r}) \delta T(\vec{r}+ \Delta \vec{r}) \rangle / \left [ \langle \delta R(\vec{r})^2 \rangle \ \langle \delta T(\vec{r})^2 \rangle \right ]^{1/2}$, which is identical to Eq.~(\ref{eq:theoryC}) when $\langle \delta R(\vec{r})^2 \rangle = \langle R(\vec{r}) \rangle^2$ and $\langle \delta T(\vec{r})^2 \rangle = \langle T(\vec{r}) \rangle^2$, i.e. when $R(\vec{r})$ and $T(\vec{r})$ obey the Rayleigh statistics (Gaussian fields). For systems close to the Anderson localization (low dimensionless conductance $g$) deviations from the Rayleigh distribution are expected~\cite{deboer92, strudley13}, but for the diffusive media under examination here the two formulas are identical to a very high accuracy (see SI section 1 for a detailed analysis).

\textbf{Numerical simulations}.
The 3D numerical simulations are performed using the coupled-dipole method~\cite{LAX-1952}. Scatterers represented by electric point dipoles are randomy distributed in a rectangular box, with longitudinal thickness $L$ and transverse size ten times larger to mimic the slab geometry used in the experiments. Since the measurements are not resolved in polarization, and since the input light is expected to depolarize on a length scale on the order of $\ell$~\cite{vynck14}, we neglect polarization and numerically solve the scalar wave equation. To limit the number of scatterers and save computational time, the polarizability $\alpha$ of each scatterer has been chosen to maximize the scattering cross-section $\sigma_s=k^4|\alpha|^2/(4\pi)$ leading to $\alpha=4i\pi/k^3$. Adjusting the number density of scatterers $\rho$, we can vary the scattering mean-free path $\ell=1/(\rho\sigma_s)$ and simulate different kinds of samples. Solving numerically the coupled-dipole equations, we compute the scattered field at any point on the input and exit surfaces of the slab, and deduce the correlation function $\langle C^{RT} (\Delta\vec{r})\rangle$. The ensemble averaging is performed by computing the field for many realizations of the positions of the scatterers. As an example, in the regime $k\ell=10$ and $b=1.5$, we have used $N=2685$ dipoles and $26$ millions of configurations.
\medskip

\textbf{Acknowledgements}\\This work was supported by the Leverhulme Trust's Philip Leverhulme Prize, and by LABEX WIFI (Laboratory of Excellence within the French Program ``Investments for the Future'') under references ANR-10-LABX-24 and ANR-10-IDEX-0001-02 PSL*. I.S. and A.M.P-D. acknowledge support from EPSRC (EP/L015331/1) through the Centre of Doctoral Training in Metamaterials (XM$^2$). N.F. acknowledges financial support from the French ``Direction G\'en\'erale de l'Armement'' (DGA).

\medskip

\textbf{Author Contributions}\\
R.C., A.G., R.P. and J.B. designed the research. I.S., A.M.P-D. and J.B. performed the experimental research. N.F., A.G., R.P. and R.C. performed the theoretical
and numerical research. All the authors contributed to writing the paper.

\medskip

\textbf{Additional Information}\\
Supplementary information is available in the online version of the paper. Reprints and permissions information is available online at www.nature.com/reprints. Correspondence and requests for materials should be addressed to J.B. (j.bertolotti@exeter.ac.uk).

\medskip

\textbf{Competing financial interests} The authors declare no competing financial interests.

\medskip

\bibliography{bibliography_2}

\clearpage

\end{document}